\documentclass[twocolumn,showpacs,prc]{revtex4-1}

\usepackage[utf8]{inputenc}
\usepackage[english]{babel}
\usepackage{tikz}
\usepackage{paralist}
\usepackage{subcaption} \usepackage{graphicx}
\usepackage{amsmath} \usepackage{amssymb}
\usepackage{xcolor}
\graphicspath{{./fig/}}

\begin{document} 

\title[Fragmentation of Neutron Star Matter]{Fragmentation of Neutron Star Matter}

\author{P. N. Alcain and C. O. Dorso}

\affiliation{Departamento de Física, FCEyN, UBA and IFIBA,
  Conicet, Pabellón 1, Ciudad Universitaria, 1428 Buenos
  Aires, Argentina} \affiliation{IFIBA-CONICET}

\date{\today} 

\begin{abstract}
\begin{description}
\item[Background] Neutron stars are astronomical systems with nucleons
  submitted to extreme conditions. Due to the long range coulomb
  repulsion between protons, the system has structural
  inhomogeneities. These structural inhomogeneities arise also in
  expanding systems, where the fragment distribution is highly
  dependent on the thermodynamic conditions (temperature, proton
  fraction, \ldots) and the expansion velocity.

\item[Purpose] We aim to find the different regimes of fragment
  distribution, and the existence of infinite clusters.
  
\item[Method] We study the dynamics of the nucleons with a
  semiclassical molecular dynamics model. Starting with an equilibrium
  configuration, we expand the system homogeneously until we arrive to
  an asymptotic configuration (i.\ e.\ very low final densities). We
  study the fragment distribution throughout this expansion.
  
\item[Results] We found the typical regimes of the asymptotic
  fragment distribution of an expansion: u-shaped, power law and
  exponential. Another key feature in our calculations is that, since
  the interaction between protons is long range repulsive, we do not
  have always an infinite fragment. We found that, as expected, the
  faster the expansion velocity is, the quicker the infinite fragment
  disappears.
  
\item[Conclusions] We have developed a novel graph-based tool for the
  identification of infinite fragments, and found a transition from
  U-shaped to exponential fragment mass distribution with increasing
  expansion rate.
\end{description}

\end{abstract}

\maketitle

\section{Introduction}\label{sc:intro}
Neutron stars appear in the universe as the remains of the death of a
massive star (but not too massive). Stars die when the internal
thermonuclear reactions in the star are no longer able to balance the
gravitational compression. At this stage, stars undergo gravitational
collapse, followed by a shock wave that ejects most of the mass of the
star leaving behind a dense core. During this collapse a big part of
electrons and protons turn into neutrons through electron capture
emitting neutrinos. Because of this, the residual mass tends to have
an excess of neutrons over protons, hence the name neutron star. It
should be emphasized that neutron stars are electrically balanced
objects: their net charge is zero.

The mass of a neutron star ranges between 1 and 3 solar masses, its
radius is of the order of a $10^{-5}$ solar radius or about 10 km,
which gives it an average density of $~10^{15}$ $g/cm^3$ or about 20
times that of normal nuclei. Energy considerations indicate that the
cores of neutron stars are topped by a \emph{crust} of about 1 km
thick, where the neutrons produced by $\beta$-decay form neutron rich
nuclear matter, immersed in a sea of electrons. Its structure can be
divided in two parts, according to current
models~\cite{page_minimal_2004, geppert_temperature_2004}: the
\emph{crust}, about 1.5 km thick and with a density of up to half the
normal nuclear density $\rho_0$; and the \emph{core}, where the
structure is still unknown and remains highly
speculative~\cite{woosley_physics_2005}.

The density of such crust ranges from normal nuclear density ($\approx
3\times10^{14}$ $g/cm^3$ or $\approx \rho_0=0.15 \text{fm}^{-3}$) at a
depth $d \approx 1 \ km$ to the neutron drip density ($\approx
4\times10^{11}$ $g/cm^3$) at $d \approx 0.5 \ km$, to, finally, an
even lighter mix of neutron rich nuclei also embedded in a sea of
electrons with densities decreasing down to practically zero in the
neutron star envelope.  Ravenhall \emph{et al.}  in
Ref.~\cite{ravenhall_structure_1983} and Hashimoto \emph{et al.} in
Ref.~\cite{hashimoto_shape_1984} proposed that the neutron star crust
is composed by the structures known as \emph{nuclear pasta}.

Several models have been developed to study nuclear pasta, and they
have shown that these structures arise due to the interplay between
nuclear and Coulomb forces in an infinite medium. Nevertheless, the
dependence of the observables on different thermodynamic quantities
has not been studied in depth. The original works of Ravenhall
\emph{et al.}~\cite{ravenhall_structure_1983} and Hashimoto \emph{et
  al.}~\cite{hashimoto_shape_1984} used a compressible liquid drop
model, and have shown that the now known as \emph{pasta phases}
--\emph{lasagna}, \emph{spaghetti} and \emph{gnocchi}-- are solutions
to the ground state of neutron star matter. From then on, different
approaches have been taken, that we roughly classify in two
categories: mean field or microscopic.

Mean field works include the Liquid Drop Model, by Lattimer \emph{et
  al.}~\cite{page_minimal_2004}, Thomas-Fermi, by Williams and
Koonin~\cite{williams_sub-saturation_1985}, among
others~\cite{oyamatsu_nuclear_1993, lorenz_neutron_1993,
  cheng_properties_1997, watanabe_thermodynamic_2000,
  watanabe_electron_2003, nakazato_gyroid_2009}. Microscopic models
include Quantum Molecular Dynamics, used by Maruyama \emph{et
  al.}~\cite{maruyama_quantum_1998, kido_md_2000} and by
Watanabe~\emph{et al.}\cite{watanabe_structure_2003}, Simple
Semiclassical Potential, by Horowitz~\emph{et
  al.}~\cite{horowitz_nonuniform_2004} and Classical Molecular
Dynamics, used in our previous works~\cite{dorso_topological_2012}.
Of the many models used to study neutron stars, the advantages of
classical or semiclassical models are the accessibility to position
and momentum of all particles at all times and the fact that no
specific shape is hardcoded in the model, as happens with most mean
field models. This allows the study of the structure of the nuclear
medium from a particle-wise point of view. Many models exist with this
goal, like quantum molecular dynamics~\cite{maruyama_quantum_1998},
simple-semiclassical potential~\cite{horowitz_nonuniform_2004} and
classical molecular dynamics~\cite{lenk_accuracy_1990}. In these
models the Pauli repulsion between nucleons of equal isospin is either
hard-coded in the interaction or as a separate
term~\cite{dorso_classical_1988}.

Semiclassical models with molecular dynamics have been used to study
the \emph{nuclear pasta} regime, with mainly two parametrizations of
the interaction: Simple Semiclassical Potential
~\cite{horowitz_nonuniform_2004}, Quantum Molecular
Dynamics~\cite{maruyama_quantum_1998} and Classical Molecular Dynamics
~\cite{dorso_topological_2012, alcain_beyond_2014}.

Multifragmentation in nuclear systems has been studied
before~\cite{bonasera_critical_2000, chikazumi_quantum_2001}, but
mostly with nuclear matter (without Coulomb interaction). In a recent
work by Caplan et al~\cite{caplan_pasta_2015}, expanding neutron star
matter has been studied as possible explanations for nucleosynthesis
in neutron star mergers.

When two neutron stars collide, a neutron star merger happens. The
compression of neutron star matter as a possible source for r-process
nuclei was first discussed in Ref.~\cite{lattimer_black-hole-neutron-star_1974}.
According to hydrodynamic models~\cite{goriely_r-process_2011}, these
have typical expansions coefficients of $\eta = 10^{-21}\,\text{c/fm} <
\eta < 4\cdot 10^{-20}\,\text{c/fm}$. Inspired by the neutron star
merger, we perform a study on the fragmentation of expanding neutron
star matter. In section~\ref{sc:nucleon} we explore different models
and, in subsection~\ref{ssc:cmd} we define the model we
use. Section~\ref{sc:expansion} explains how we simulate the expansion
of the system. To analyze fragments, we describe in
section~\ref{sc:cluster} the cluster recognition algorithm, with
emphasis on the identification of infinite fragments.

\section{Nucleon Dynamics}\label{sc:nucleon}
To study the nuclear structure of stellar crusts is necessary
understand the behavior of nucleons, i.e. protons and neutrons, at
densities and temperatures as those encountered in the outer
layers of stars.  This knowledge comes from decades of study of
nucleon dynamics through the use of computational models that have
evolved from simple to complex.  In this section we review this
evolution, presenting a synopsis of existing models and culminating
with the model used in this work, namely classical molecular
dynamics (CMD).

\subsection{Evolution of Models}
Several approaches have been used to simulate the behavior of nuclear
matter, especially in reactions. The initial statistical studies of
the 1980's, which lacked fragment-fragment interactions and
after-breakup fragment dynamics~\cite{barz_cluster_1996}, were
followed in the early 1990s by more robust studies that incorporated
these effects.  All these approaches, however, were based on idealized
geometries and missed important shape fluctuation and
reaction-kinematic effects.  On the transport-theory side, models were
developed in the late 1990s based on classical, semiclassical, and
quantum-models; because of their use in the study of stellar nuclear
systems nowadays, a side-by-side comparison of these transport models
is in order.

Starting with the semiclassical, a class of models known generically
as BUU is based on the Vlasov-Nordheim
equation~\cite{nordheim_kinetic_1928} also known as the
Boltzmann-Uehling-Uhlenbeck~\cite{uehling_transport_1933}. These
models numerically track the time evolution of the Wigner function,
$f(\mathbf{r},\mathbf{p})$, under a mean potential $U(\mathbf{r})$ to
obtain a description of the probability of finding a particle at a
point in phase space. The semiclassical interpretation constitutes
BUU's main advantage; the disadvantages come from the limitation of
using only a mean field which does not lead to cluster formation. To
produce fragments, fluctuations must be added by hand, and more
add-ons are necessary for secondary decays and other more realistic
features.

On the quantum side, the molecular dynamics models, known as QMD,
generically speaking, solve the equations of motion of nucleon
wave packets moving within mean fields (derived from Skyrme potential
energy density functional).  The method allows the imposition of a
Pauli-like blocking mechanism, use of isospin-dependent
nucleon-nucleon cross sections, momentum dependence interactions and
other variations to satisfy the operator's tastes.  The main advantage
is that QMD is capable of producing fragments, but at the cost of a
poor description of cluster properties and the need of sequential
decay codes external to BUU to cut the fragments and de-excite them
\cite{polanski_development_2005}; normally, clusters are constructed
by a coalescence model based on distances and relative momenta of
pairs of nucleons. Variations of these models applied to stellar
crusts can be found elsewhere in this volume.

Of particular importance to this article is the classical molecular
dynamics (CMD) model, brainchild of the Urbana group
~\cite{lenk_accuracy_1990} and designed to reproduce the predictions
of the Vlasov-Nordheim equation while providing a more complete
description of heavy ion reactions.  As it will be described in more
detail in the next section, the model is based on the Pandharipande
potential which provides the ``nuclear'' interaction through a
combination of Yukawa potentials selected to correspond to infinite
nuclear matter with proper equilibrium density, energy per particle,
and compressibility.

Problems common to both BUU and QMD are the failure to produce
appropriate number of clusters, and the use of hidden adjustable
parameters, such as the width of wave packets, number of test
particles, modifications of mean fields, effective masses and cross
sections.  These problems are not present in the classical molecular
dynamics model, which without any adjustable parameters, hidden or
not, is able to describe the dynamics of the reaction in space from
beginning to end and with proper energy, space, and time units.  It
intrinsically includes all particle correlations at all levels:
2-body, 3-body, etc. and can describe nuclear systems ranging from
highly correlated cold nuclei (such as two approaching heavy ions), to
hot and dense nuclear matter (nuclei fused into an excited blob), to
phase transitions (fragment and light particle production), to
hydrodynamics flow (after-breakup expansion) and secondary decays
(nucleon and light particle emission).

The only apparent disadvantage of the CMD is the lack of quantum
effects, such as the Pauli blocking, which at medium excitation
energies stops the method from describing nuclear structure
correctly. Fortunately, in collisions, the large energy deposition
opens widely the phase space available for nucleons and renders Pauli
blocking practically obsolete~\cite{lopez_lectures_2000}, while in
stellar environments, at extremely low energies and with frozen-like
structures, momentum-transferring collisions cease to be an important
factor in deciding the stable configuration of the nuclear
matter. Independent of that, the role of quantum effects in two body
collisions is guaranteed to be included by the effectiveness of the
potential in reproducing the proper cross sections, furthermore, an
alternate fix is the use of momentum dependent potentials (as
introduced by Dorso and Randrup~\cite{dorso_classical_1988}) when
needed.

Although no theories can yet claim to be a perfect description of
nuclear matter, all approaches have their advantages and drawbacks
and, if anything can be said about them, is that they appear to be
complementary to each other.

\subsection{Classical Molecular Dynamics}\label{ssc:cmd}
In this work, we study fragmentation of Neutron Star Matter under
pasta-like conditions with the classical molecular dynamics model
CMD.\@ It has been used in several heavy-ion reaction studies to: help
understand experimental data~\cite{chernomoretz_quasiclassical_2002};
identify phase-transition signals and other critical
phenomena~\cite{lopez_lectures_2000, barranon_searching_2001,
  dorso_selection_2001, barranon_critical_2003, barranon_time_2007};
and explore the caloric curve~\cite{barranon_entropy_2004} and
isoscaling~\cite{dorso_dynamical_2006, dorso_isoscaling_2011}. CMD
uses two two-body potentials to describe the interaction of nucleons,
which are a combination of Yukawa potentials:
\begin{align*}
  V^{\text{CMD}}_{np}(r) &=v_{r}\exp(-\mu_{r}r)/{r}-v_{a}\exp(-\mu_{a}r)/{r}\\ 
  V^{\text{CMD}}_{nn}(r) &=v_{0}\exp(-\mu_{0}r)/{r}
\end{align*}
where $V_{np}$ is the potential between a neutron and a proton, and
$V_{nn}$ is the repulsive interaction between either $nn$ or $pp$. The
cutoff radius is $r_c=5.4\,\text{fm}$ and for $r>r_c$ both potentials
are set to zero. The Yukawa parameters $\mu_r$, $\mu_a$ and $\mu_0$
were determined to yield an equilibrium density of $\rho_0=0.16
\,\text{fm}^{-3}$, a binding energy $E(\rho_0)=16
\,\text{MeV/nucleon}$ and a compressibility of $250\,\text{MeV}$.

To simulate an infinite medium, we used this potential with
$N = 11000$ particles under periodic boundary conditions, with
different proton fraction (i.e. with $x = Z/A = 0.2 < x < 0.4$) in
cubical boxes with sizes adjusted to have densities
$\rho=0.05 \,\text{fm}^{-3}$ and $\rho = 0.08\,\text{fm}^{-3}$. These
simulations have been done with LAMMPS~\cite{plimpton_fast_1995},
using its GPU package~\cite{brown_implementing_2012}.

\subsubsection{Ground State Nuclei} 
\begin{figure}[h]
  \centering
  \includegraphics[width=0.8\columnwidth]{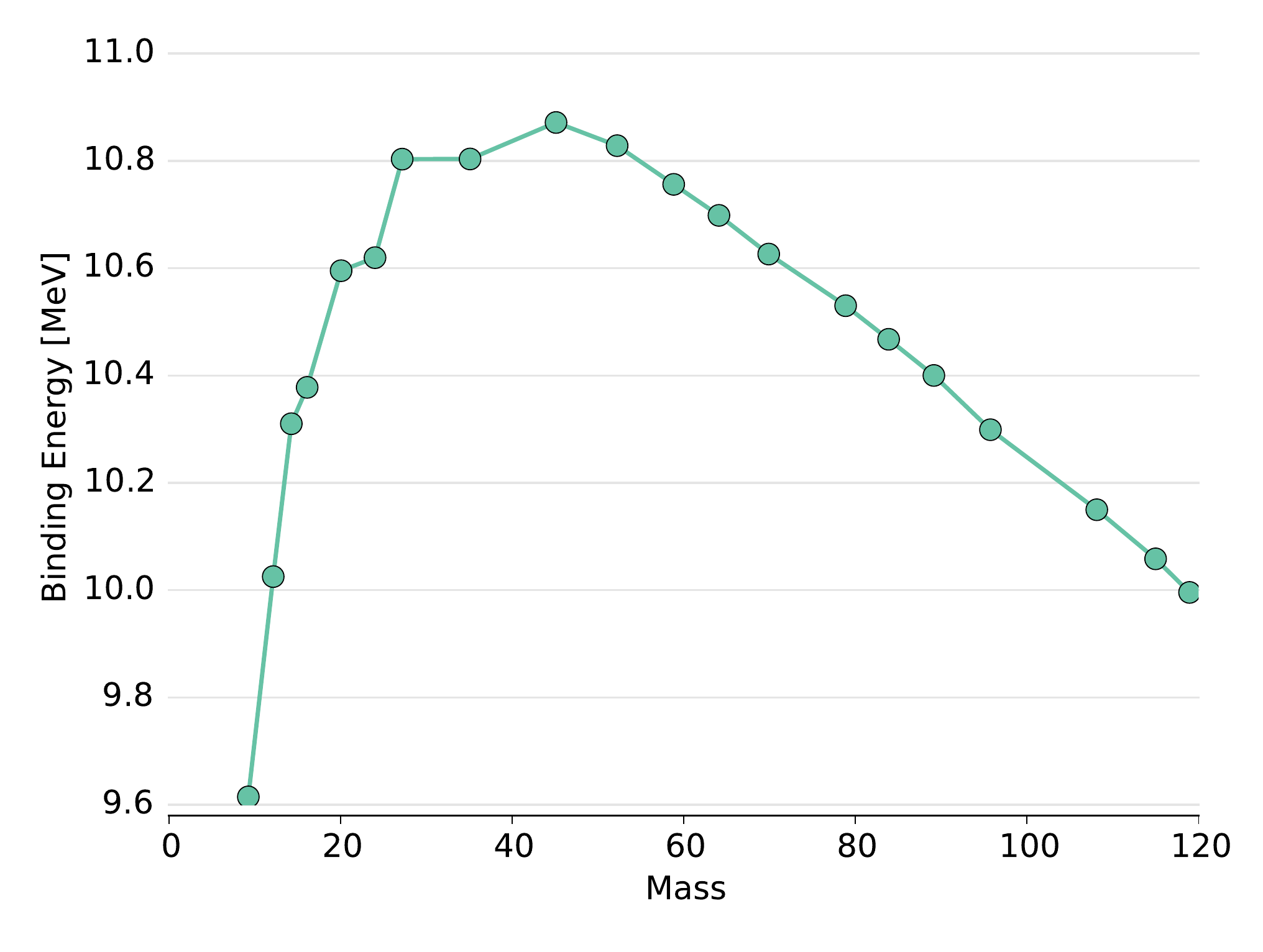}
  \caption{Binding energies of ground-state nuclei obtained with CMD
    model, extracted from Ref.~\cite{dorso_isoscaling_2011}.}
  \label{fig:binding}
\end{figure}

Although the $T=0$ state of this classical nuclear matter at normal
densities is a simple cubic solid, nuclear systems can be mimicked by
adding enough kinetic energy to the nucleons. To study nuclei, for
instance, liquid-like spherical drops with the right number of protons
and neutrons are constructed confined in a steep spherical potential
and then brought to the ``ground'' state by cooling them slowly from a
rather high temperature until they reach a self-contained state.
Removing the confining potential, the system is further cooled down
until a reasonable binding energy is attained. The remaining kinetic
energy of the nucleons helps to resemble the Fermi motion.
Figure~\ref{fig:binding} shows the binding energies of ``ground-state
nuclei'' obtained with the mass formula and with CMD;
see~\cite{dorso_isoscaling_2011} for details.

\subsubsection{Collisions} 
With respect to collisions, these potentials are known to reproduce
nucleon-nucleon cross sections from low to intermediate
energies~\cite{lenk_accuracy_1990}, and it has been used extensively
in studying heavy ion collisions (see
Ref.~\cite{chernomoretz_quasiclassical_2002, barranon_time_2007}). For
such reactions, two ``nuclei'' are boosted against each other at a
desired energy. From collision to collision, the projectile and target
are rotated with respect to each other at random values of the Euler
angles. The evolution of the system is followed using a
velocity-Verlet algorithm with energy conservation better than 0.01\%.
At any point in time, the nucleon information, i.e., position and
momenta, can be turned into fragment information by identifying the
clusters and free particles; several such cluster recognition
algorithms have been developed by our collaborator, C.O. Dorso, and
they are well described in the
literature~\cite{dorso_when_1995,strachan_time_1997}.

The method yields mass multiplicities, momenta, excitation energies,
secondary decay yields, etc. comparable to experimental
data~\cite{belkacem_searching_1996,chernomoretz_quasiclassical_2002}.
Figure~\ref{fig:distribution}, for instance, shows experimental and
simulated parallel velocity distributions for particles obtained from
mid-peripheral and peripheral ${}^{58}Ni+C$ collisions performed at
the Coupled Tandem and Super-Conducting Cyclotron accelerators of AECL
at Chalk River~\cite{chernomoretz_quasiclassical_2002}.

\begin{figure}[h]
  \begin{subfigure}[h!]{\columnwidth}
    \includegraphics[width=\columnwidth]{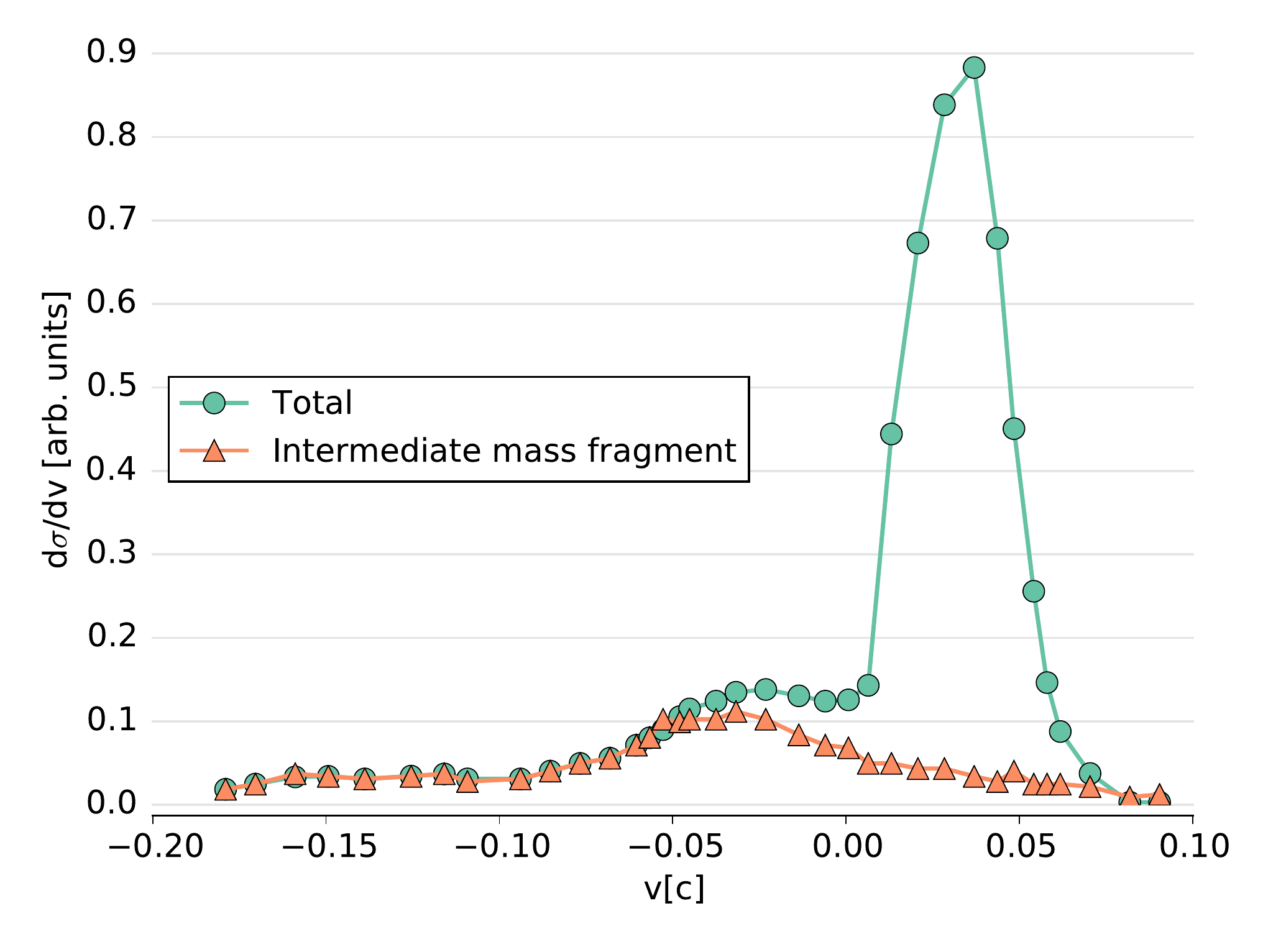}
    \caption{$\eta = 0.0001\,\text{fm/c}$}
    \label{sfig:exp}
  \end{subfigure}
  \begin{subfigure}[h!]{\columnwidth}
    \includegraphics[width=\columnwidth]{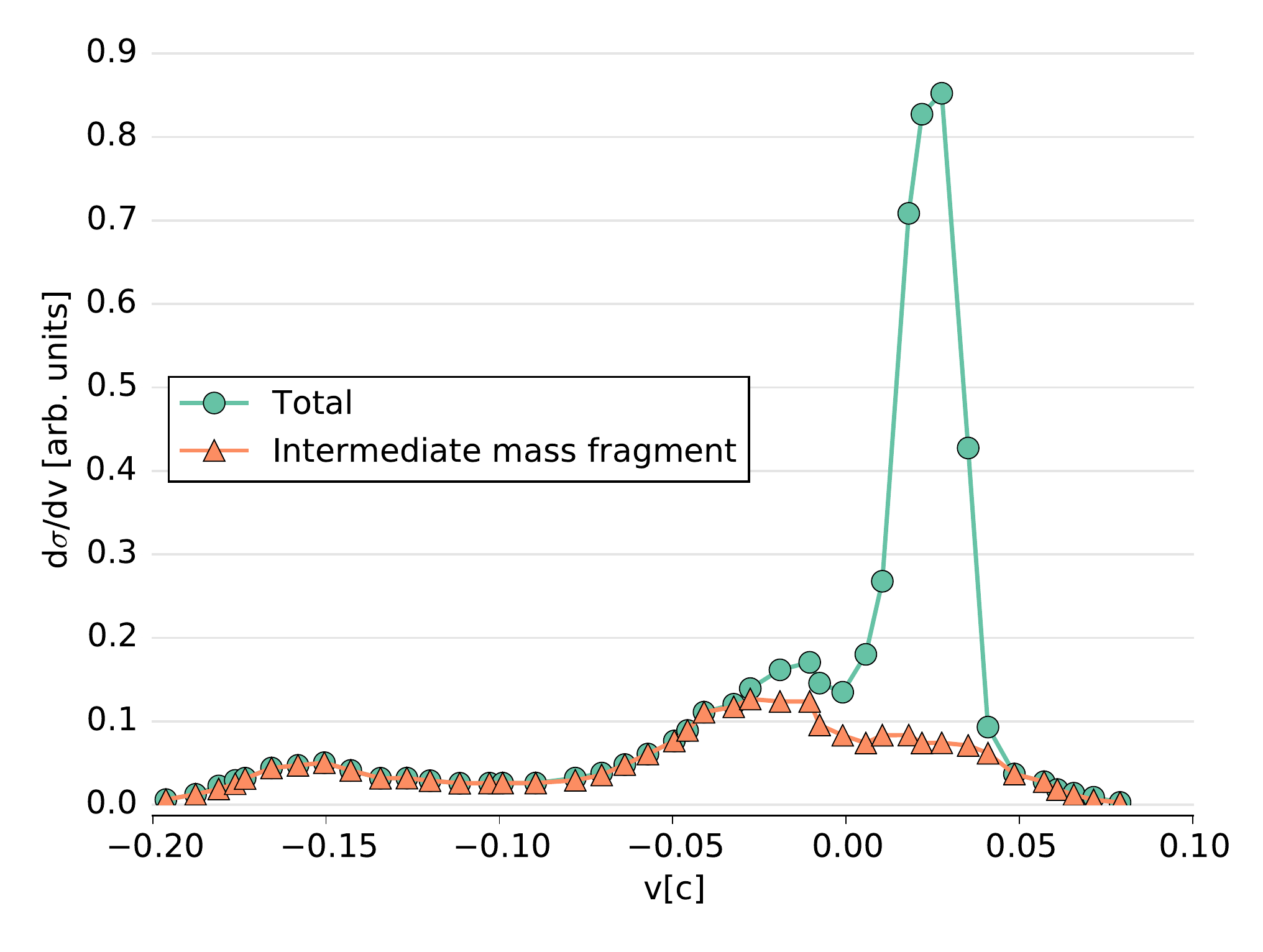}
    \caption{$\eta = 0.0001\,\text{fm/c}$}
    \label{sfig:sim}
  \end{subfigure}
  \centering
  \caption{\ref{sfig:exp} Experimental and \ref{sfig:sim} simulated
    parallel velocity distributions for ${}^{58}Ni+C$ collisions.}
  \label{fig:distribution}
\end{figure}

\subsubsection{Thermostatic Properties of Nuclear Matter} To study
thermal properties of static nuclear matter, these drops or infinite
systems, nucleons are positioned at random, but with a selected
density, in a container and ``heated". After equilibration, the system
can then be used to extract macroscopic variables. Repeating these
simulations for a wide range of density and temperature values,
information about the energy per nucleon, $\epsilon(\rho,T)$, can be
obtained and used to construct analytical fits in the spirit of those
pioneered by Bertsch, Siemens, and Kapusta~\cite{bertsch_nuclear_1983,
  kapusta_deuteron_1984, lopez_nuclear_1984}; these fits in turn can
be used to derive other thermodynamic variables, such as pressure,
etc.

\begin{figure}[h]
  \centering
  \includegraphics[width=\columnwidth]{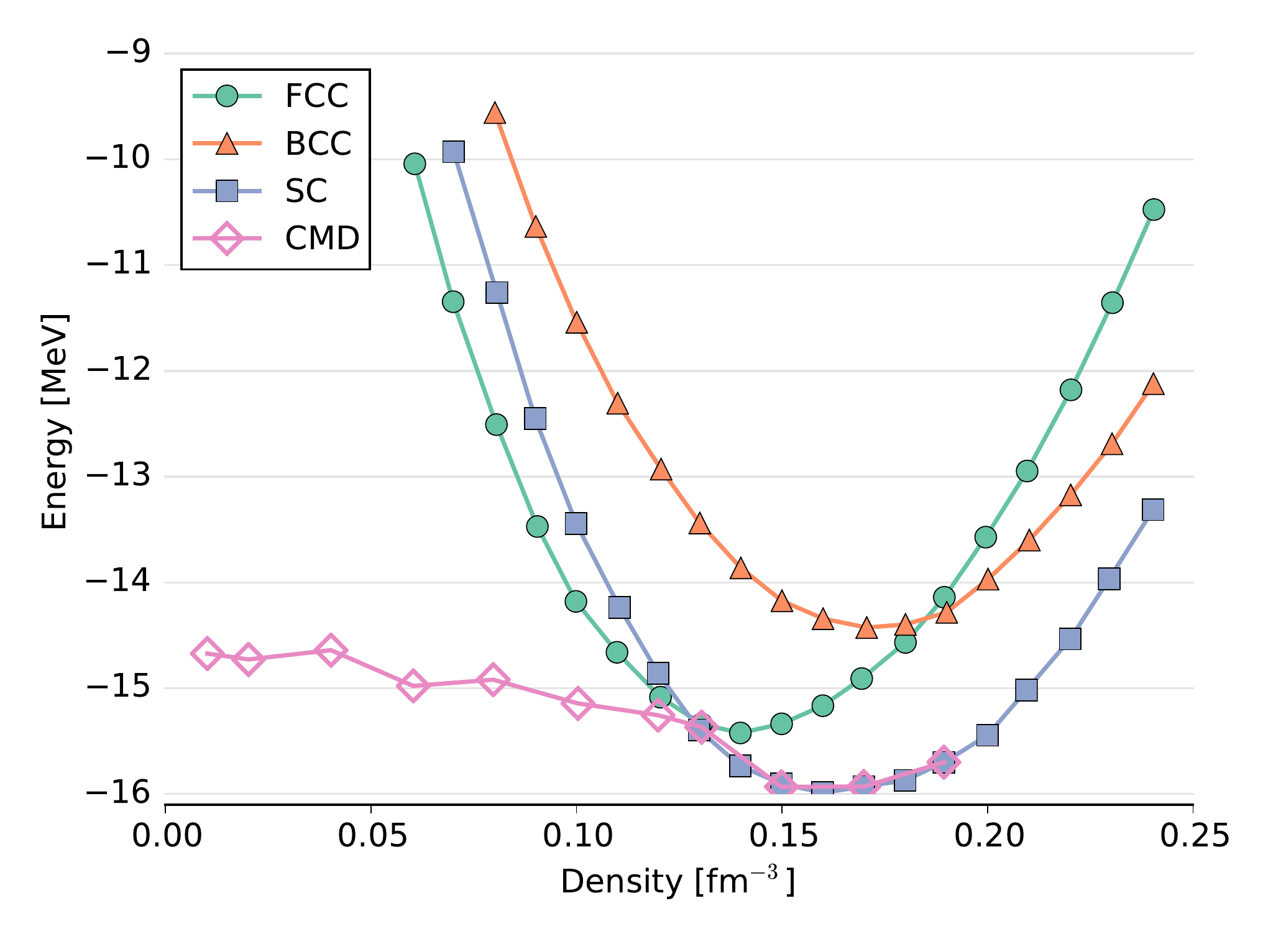}
  \caption{Nuclear matter energy per particle of cold matter
    calculated with CMD. Full symbols are for homogeneous systems,
    while open circles denote the emergence of the pseudo-pasta.}
  \label{fig:energy_nm}
\end{figure}

Figure~\ref{fig:energy_nm} shows the results of the method as applied
by Gim\'enez-Molinelli \emph{et
  al.}~\cite{gimenez_molinelli_simulations_2014} for stiff cold
infinite nuclear matter. The open symbols correspond to CMD
calculations at low temperatures, showing a departure from the imposed
homogeneous solutions obtained with the full symbols for different
crystal symmetries. This shows the emergence of pseudo pastas in
nuclear matter.

\begin{figure}[h]
  \centering
  \includegraphics[width=\columnwidth]{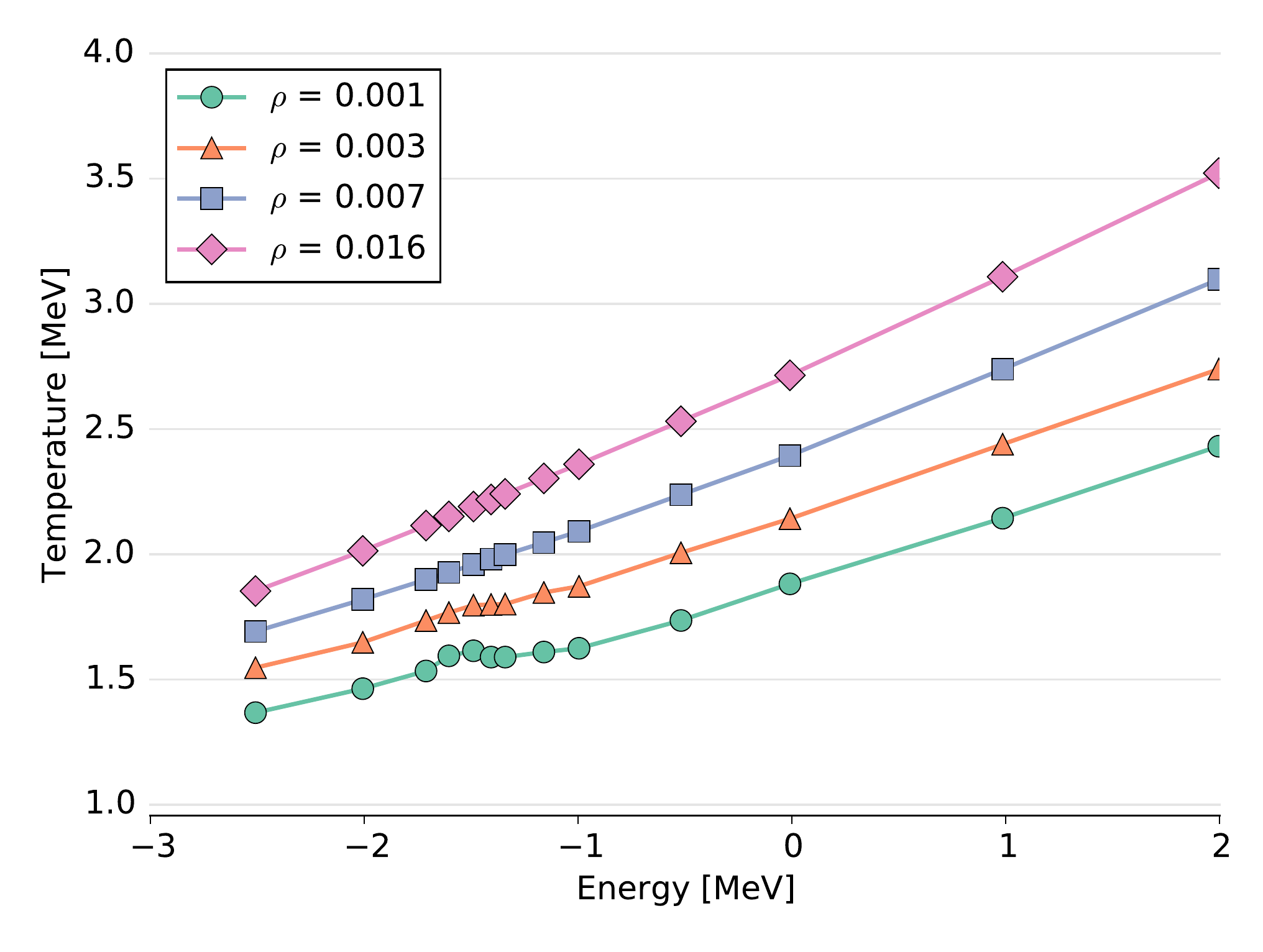}
  \caption{Caloric curve of a system equilibrated at four different
    densities calculated with CMD.}
  \label{fig:caloric}
\end{figure}

For finite systems, figure~\ref{fig:caloric} demonstrates the
feasibility of using CMD to study thermal properties such as the
caloric curve (i.e. the temperature - excitation energy relationship)
for a system of 80 nucleons equilibrated at four different densities;
see Ref.~\cite{dorso_isoscaling_2011} for complete details.

\subsection{Coulomb interaction in the model}\label{sc:coulomb}

Since a neutralizing electron gas embeds the nucleons in the neutron
star crust, the Coulomb forces among protons are screened. The model
we used to model this screening effect is the Thomas-Fermi
approximation, used with various nuclear
models~\cite{maruyama_quantum_1998, dorso_topological_2012,
  horowitz_neutrino-pasta_2004}. According to this approximation,
protons interact via a Yukawa-like potential, with a screening length
$\lambda$:
\begin{equation*}
 V_{TF}(r) = q^2\frac{e^{-r/\lambda}}{r}.
\end{equation*}

Theoretical estimates for the screening length $\lambda$ are
$\lambda\sim100\,\text{fm}$~\cite{fetter_quantum_2003}, but we set the
screening length to $\lambda=20\,\text{fm}$. This choice was based on
previous studies~\cite{alcain_effect_2014}, where we have shown that
this value is enough to adequately reproduce the expected length scale
of density fluctuations for this model, while larger screening lengths
would be a computational difficulty. We analyze the opacity to
neutrinos of the structures for different proton fractions and
densities.

\section{Neutron Star Matter at low densities and
  temperature}\label{sc:nsm_lowd}

\begin{figure}[h]
  \begin{subfigure}[h!]{0.45\columnwidth}
    \includegraphics[width=\columnwidth]{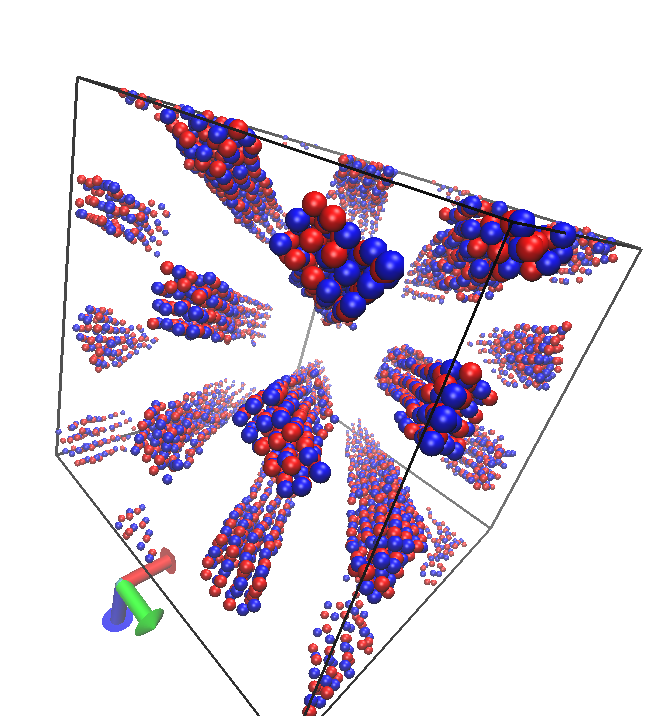}
    \caption{$\rho = 0.03\,\text{fm}^{-3}$}
    \label{sfig:spaghetti}
  \end{subfigure}
  \begin{subfigure}[h!]{0.45\columnwidth}
    \includegraphics[width=\columnwidth]{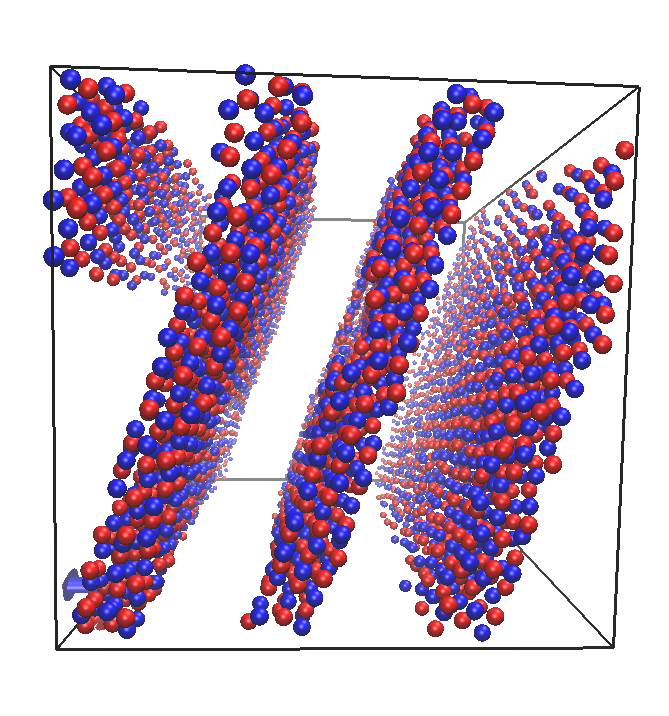}
    \caption{$\rho = 0.05\,\text{fm}^{-3}$}
    \label{sfig:lasagna}
  \end{subfigure}
  \begin{subfigure}[h!]{0.45\columnwidth}
    \includegraphics[width=\columnwidth]{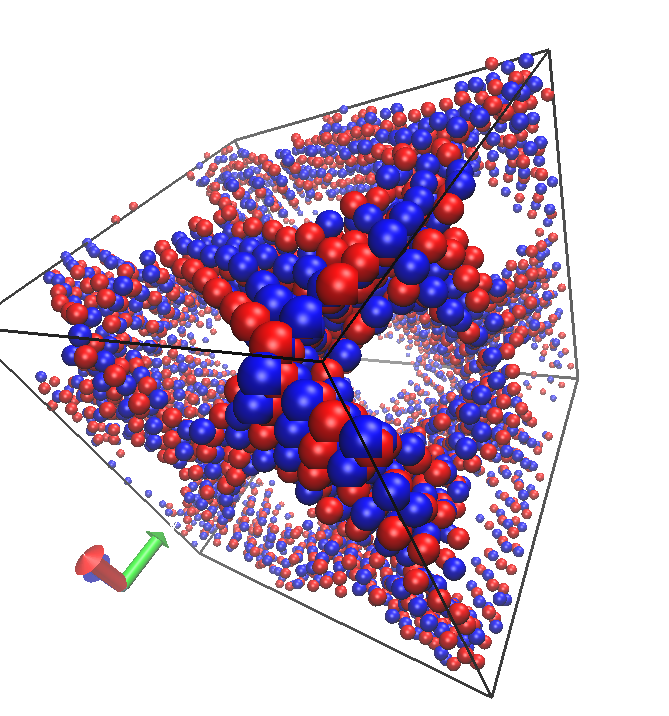}
    \caption{$\rho = 0.08\,\text{fm}^{-3}$}
    \label{sfig:tunnels}
  \end{subfigure}
  \centering
  \caption{\ref{sfig:spaghetti} Spaghetti, \ref{sfig:lasagna} lasagna
    and \ref{sfig:tunnels} tunnels obtained with the CMD method, for
    the symmetric case $x=0.5$ and low temperature $T=0.5\,\text{MeV}$
    extracted from Ref.~\cite{alcain_effect_2014}}
  \label{fig:pasta}
\end{figure}

When we consider the system with a screened Coulomb interaction as
described in~\ref{sc:coulomb}, a very interesting phenomena takes
place, described with detail in Ref.~\cite{alcain_effect_2014}. At
sub-saturation densities, the system displays an inhomogeneous
structure known as nuclear pasta, characterized by the emergence of
multiple structure per simulation cell, that can be roughly classified
as \emph{gnocchi}, \emph{spaghetti}, \emph{lasagna} and tunnels. As an
example, in figure~\ref{fig:pasta} we show the configurations of some
of these pastas.

\section{Expansion}\label{sc:expansion}

To simulate an expanding system we scale linearly with time the length
of the box in every dimension,
\begin{equation*}
  L(t) = L_0 (1 + \eta\,t)
\end{equation*}
This, however, is not enough to expand the system collectively. We
also need the particles inside the cell to expand like the box. In
order to accomplish this, based on Ref.~\cite{dorso_onset_1996}, we
add to each particle a velocity $v_{exp}$ dependent on the position in
the box:
\begin{equation*}
  \mathbf{v_{exp}} = \eta\,\mathbf{r}
\end{equation*}
We can see from this expression that the particles in the edge of the
box will have an expanding velocity equal to that of the box.

Another effect to consider of this expansion is that when a particle
crosses a boundary its velocity has to change according to the
velocity of the expanding box. For example, if the particle crosses
the left-hand boundary of the periodic box, the velocity of the image
particle $v_i^\dagger$ on the right-hand must be modified $v_i^\dagger
= v_i + L_0\,\eta$.

\section{Cluster recognition}\label{sc:cluster}
In typical configurations we have not only the structure known as
nuclear pasta, but also a nucleon gas that surrounds the nuclear
pasta. In order to properly characterize the pasta phases, we must
know which particles belong to the pasta phases and which belong to
this gas. To do so, we have to find the clusters that are formed along
the simulation.

One of the algorithms to identify cluster formation is Minimum
Spanning Tree (MST). In MST algorithm, two particles belong to the
same cluster $\{C^{\text{MST}}_n\}$ if the relative distance of the
particles is less than a cutoff distance $r_{cut}$:
\begin{equation*}
  i \in C^{\text{MST}}_n \Leftrightarrow \exists j \in C_n \mid
  r_{ij} < r_{cut}
\end{equation*}

This cluster definition works correctly for systems with no kinetic
energy, and it is based in the attractive tail of the nuclear
interaction. However, if the particles have a non-zero temperature, we
can have a situation of two particles that are closer than the cutoff
radius, but with a large relative kinetic energy. 

To deal with situations of non-zero temperatures, we need to take into
account the relative momentum among particles. One of the most
sophisticated methods to accomplish this is the Early Cluster
Recognition Algorithm (ECRA)~\cite{dorso_early_1993}. In this
algorithm, the particles are partitioned in different disjoint
clusters $C^{\text{ECRA}}_n$, with the total energy in each cluster:
\begin{equation*}
  \epsilon_n = \sum_{i \in C_n} K^{CM}_i +  \sum_{i,j \in C_n} V_{ij}
\end{equation*}
where $K^{CM}_i$ is the kinetic energy relative to the center of mass
of the cluster. The set of clusters $\{C_n\}$ then is the one that
minimizes the sum of all the cluster energies $E_{\text{partition}} =
\sum_n \epsilon_n$.

ECRA algorithm can be easily used for small
systems~\cite{dorso_fluctuation_1994}, but being a combinatorial
optimization, it cannot be used in large systems. While finding ECRA
clusters is very expensive computationally, using simply MST clusters
can give extremely biased results towards large clusters. We have
decided to go for a middle ground choice, the Minimum Spanning Tree
Energy (MSTE) algorithm~\cite{dorso_topological_2012}. This algorithm
is a modification of MST, taking into account the kinetic
energy. According to MSTE, two particles belong to the same cluster
$\{C^{\text{MSTE}}_n\}$ if they are energy bound:
\begin{equation*}
  i \in C^{\text{MSTE}}_n \Leftrightarrow \exists j \in C_n :
  V_{ij}+ K_{ij} \le 0
\end{equation*}
While this algorithm doesn't yield the same theoretically sound
results from ECRA, it still avoids the largest pitfall of naïve MST
implementations for the temperatures used in this work.

\subsection{Infinite Clusters}
\begin{figure}  \centering
  \includegraphics[width=0.75\columnwidth]{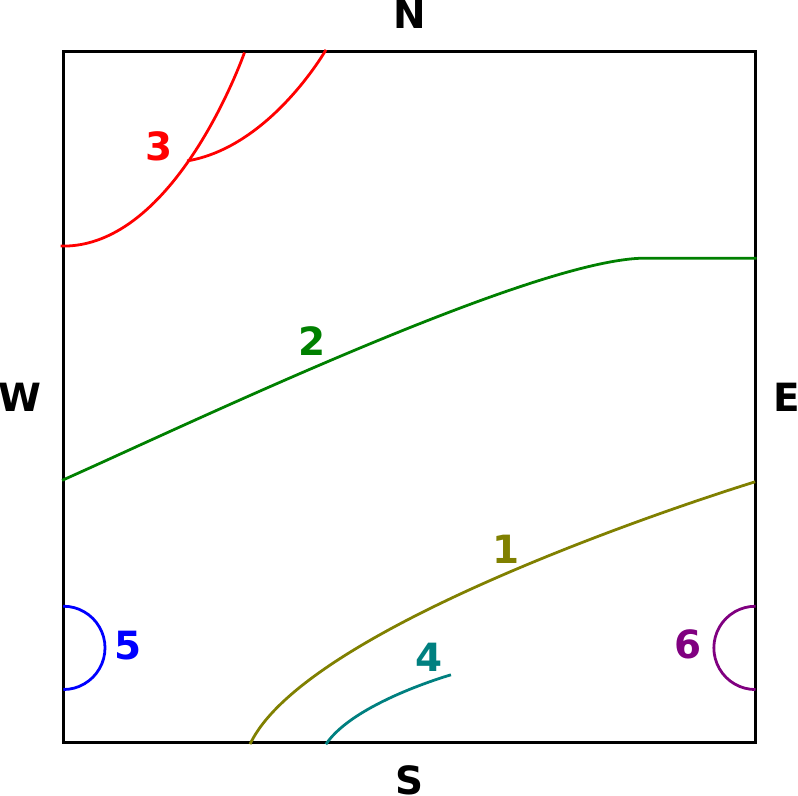}
  \caption{(Color online) Schematical representation of 2D clusters, recognized only
    in the cell and not through the periodic walls, labeled as N, S,
    W, E. The clusters inside the cell are labeled from 1 to 6,}
\label{fig:scheme_clusters}
\end{figure}

We developed an algorithm for the recognition of infinite clusters
across the boundaries. We explain here in detail the implementation
for MST clusters in 2D, being the MSTE and 3D extension
straightforward. In figure~\ref{fig:scheme_clusters} we see a
schematical representation of 2D clusters recognized in a periodic
cell, labeled from 1 to 6 (note that these clusters don't connect yet
through the periodic walls).

\begin{figure}  \centering
  \includegraphics[width=0.45\columnwidth]{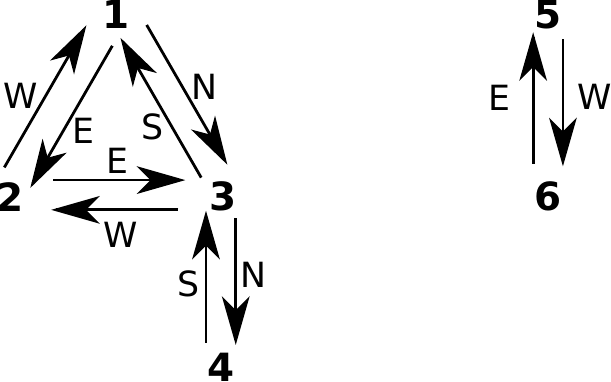}
  \caption{Graph of the clusters with connections labeled by the wall
    of the boundary they connect through. The graph can be divided in
    2 subgraphs that don't connect: 1--2--3--4 and 5--6. Each of these
    subgraphs is as cluster when periodic boundary conditions are
    considered.}
\label{fig:graph_clusters}
\end{figure}

In order to find the connections of these clusters through the
boundaries, we draw a labeled graph of the clusters, where we connect
clusters depending on whether they connect or not through a wall and
label such connection with the wall label. For example, we begin with
cluster 1. It connects with cluster 2 going out through the E wall,
therefore we add a $1\rightarrow2$ connection labeled as
E. Symmetrically, we add a $2\rightarrow1$ connection labeled as W. Now
we go for the pair 1-3. It connects going out through the S wall, so
we add $1\rightarrow3$ labeled as S and $3\rightarrow1$ labeled as
N. Cluster 1 does not connect with 4, 5, or 6, therefore those are the
only connections we have. Once we've done that, we get the graph of
figure~\ref{fig:graph_clusters}.

We now wonder whether these subgraphs represent an infinite cluster or
not not. In order to have an infinite clusters, we need to have a loop
(the opposite is not true: having a loop is not enough to have an
infinite cluster, as we can see in subragph 5--6), so we first
identify loops and mark them as candidates for infinite
clusters. Every connection adds to a loop (since the graph connections
are back and forth), but we know from inspecting the
figure~\ref{fig:graph_clusters} that the cluster 1--2--3 is
infinite. Finding out what makes, in the graph, the cluster 1--2--3
infinite is key to identify infinite clusters. And the key feature of
cluster 1--2--3 is that its loop 1--2--3--1 can be transversed through
the walls E--E--S, while loops like 5--6 can be transversed only
through E--W. Now, in order for the cluster to be infinite, we need it
to extend infinitely in (at least) one direction. So once we have the
list of walls of the loop, we create a magnitude I associated to each
loop that is created as follows: beginning with $I = 0$, we add a
value $M_i$ if there is (at least one) $i$ wall. The values are: $M_E
= 1$, $M_W = -1$, $M_N = 2$, $M_S = -2$. If $I$ is nonzero, then the
loop is infinite. For example, for the loop E--E--S, we have E and S
walls, so $I = M_E + M_S = 3$ and the loop is infinite. For the loop
E--W, $I = M_E + M_W = 0$, and the loop is finite.

\section{Results}

\begin{figure} \centering
  \begin{subfigure}[h!]{0.45\columnwidth}
    \includegraphics[width=\columnwidth]{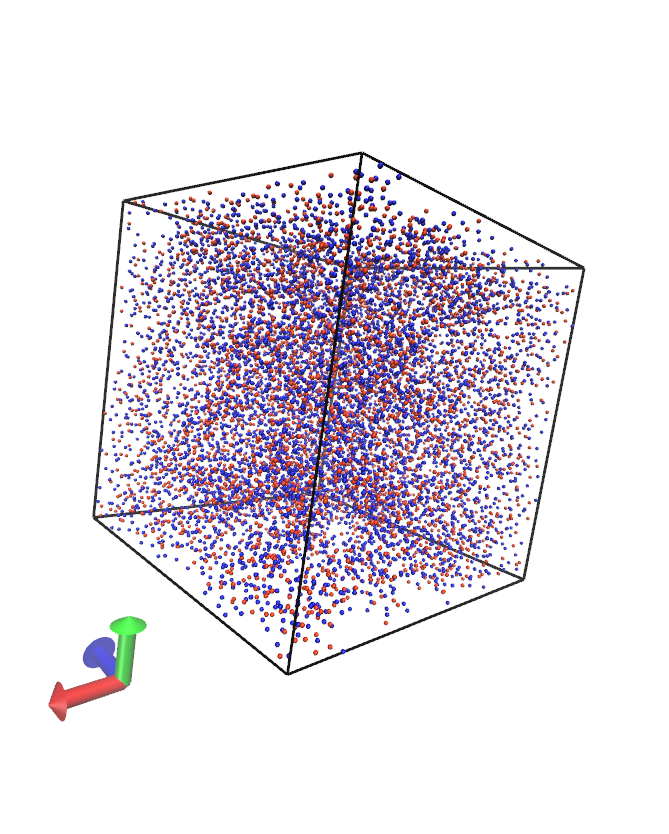}
    \caption{$\eta = 0.0001\,\text{fm/c}$}
    \label{subfig:initial}
  \end{subfigure}
  \begin{subfigure}[h!]{0.45\columnwidth}
    \includegraphics[width=\columnwidth]{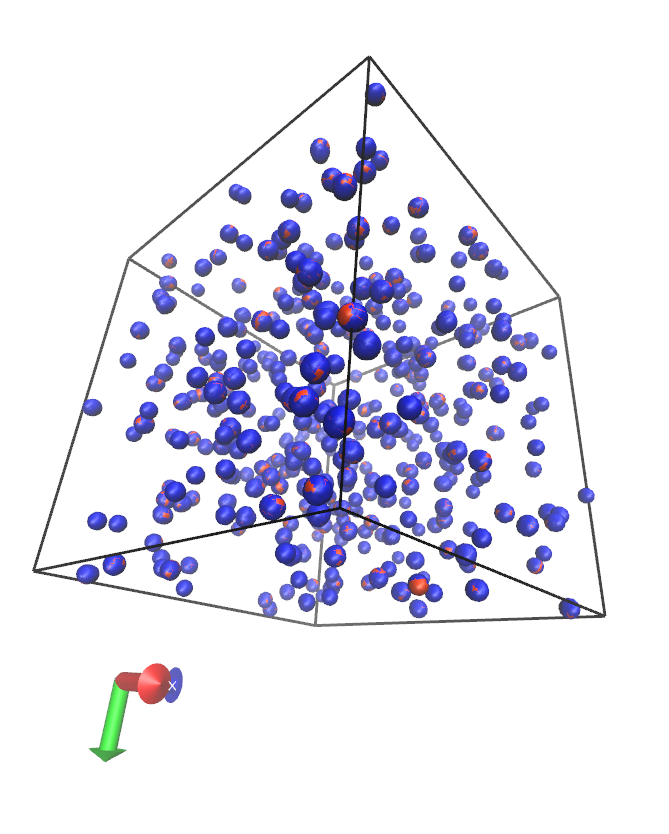}
    \caption{$\eta = 0.0005\,\text{fm/c}$}
    \label{subfig:expanded}
  \end{subfigure}
  \caption{(Color online) Snapshots of a system in the initial
    configuration~\ref{subfig:initial} and the final expanded
    system~\ref{subfig:expanded}. In the expanded system we see
    clearly formed \emph{gnocchi} clusters}
  \label{fig:morpho}
\end{figure}

In figure~\ref{fig:morpho} we show the initial and final states for
the expansion of $N=11000$ particle primordial cell (see caption for
details) for the case of low velocity expansion. It can be immediately
seen that though the initial configuration shows a compact particle
distribution, the final configuration consists of \emph{gnocchi}
structures (almost spherical fragments) with a mass of about 80
particles.

\begin{figure}
  \centering
  \includegraphics[width=\columnwidth]{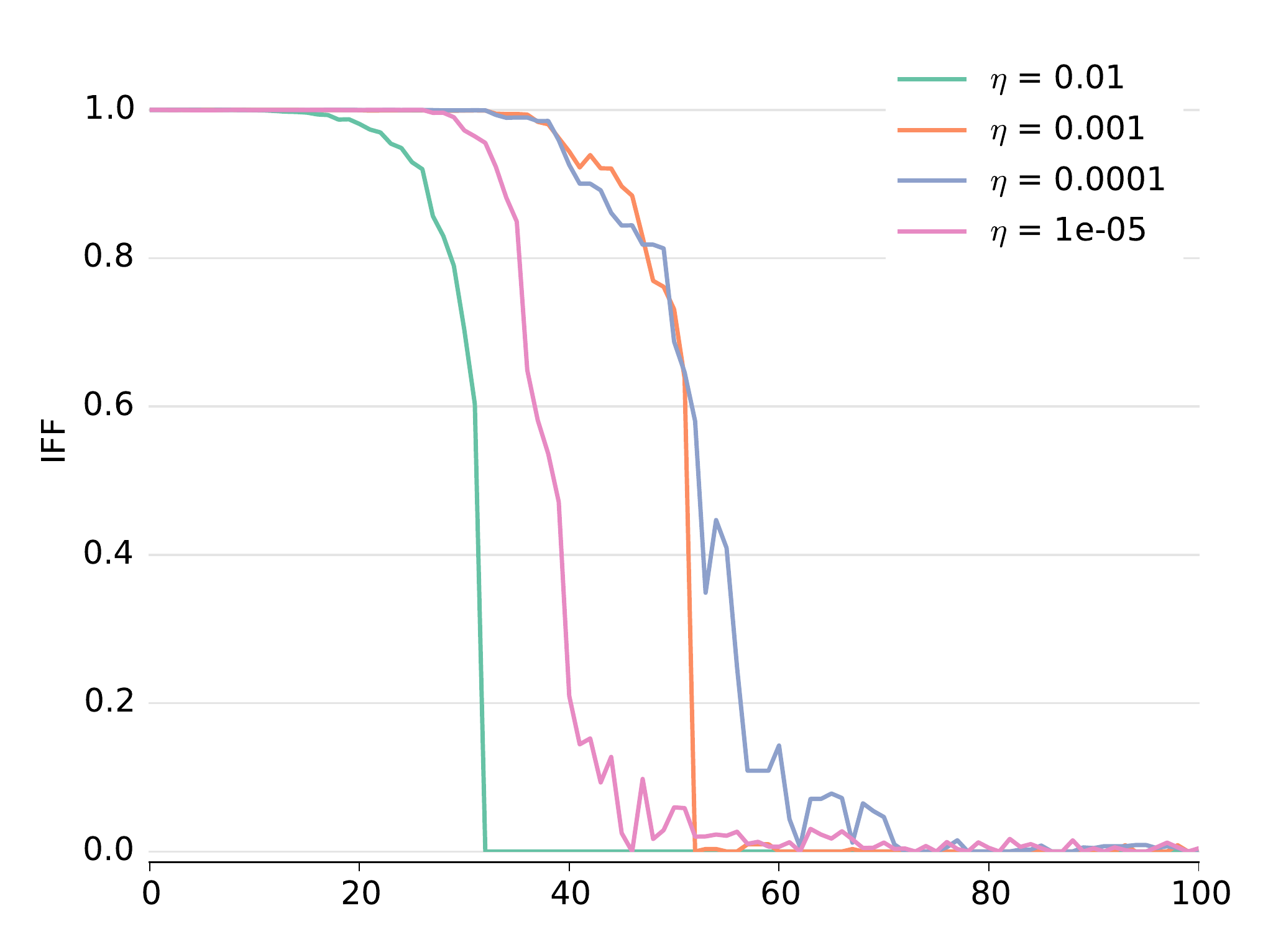}
  \caption{(Color online) Infinite Fragment Fraction (see text) as a
    function of the length of the edge of the primordial cell. For
    every expansion rate displayed the IFF goes to zero in the
    asymptotic regime.}
  \label{fig:infinite}
\end{figure}

In figure~\ref{fig:infinite} we show the fraction of particles in the
primordial cell that form part of an infinite cluster (\emph{Infinite
  Fragment Fraction}, IFF). It can be easily seen that in the early
stages of the evolution, due to the fact that the temperature is low,
most of the system in the primordial cell belongs to the infinite
clusters. But as the system evolves according to the expansion rate as
explained above, the IFF goes down and goes to zero rather quickly,
meaning that there is no more infinite fragment in the system. See
caption for details.

\begin{figure*} \centering
  \begin{subfigure}[h!]{0.9\columnwidth}
    \includegraphics[width=\columnwidth]{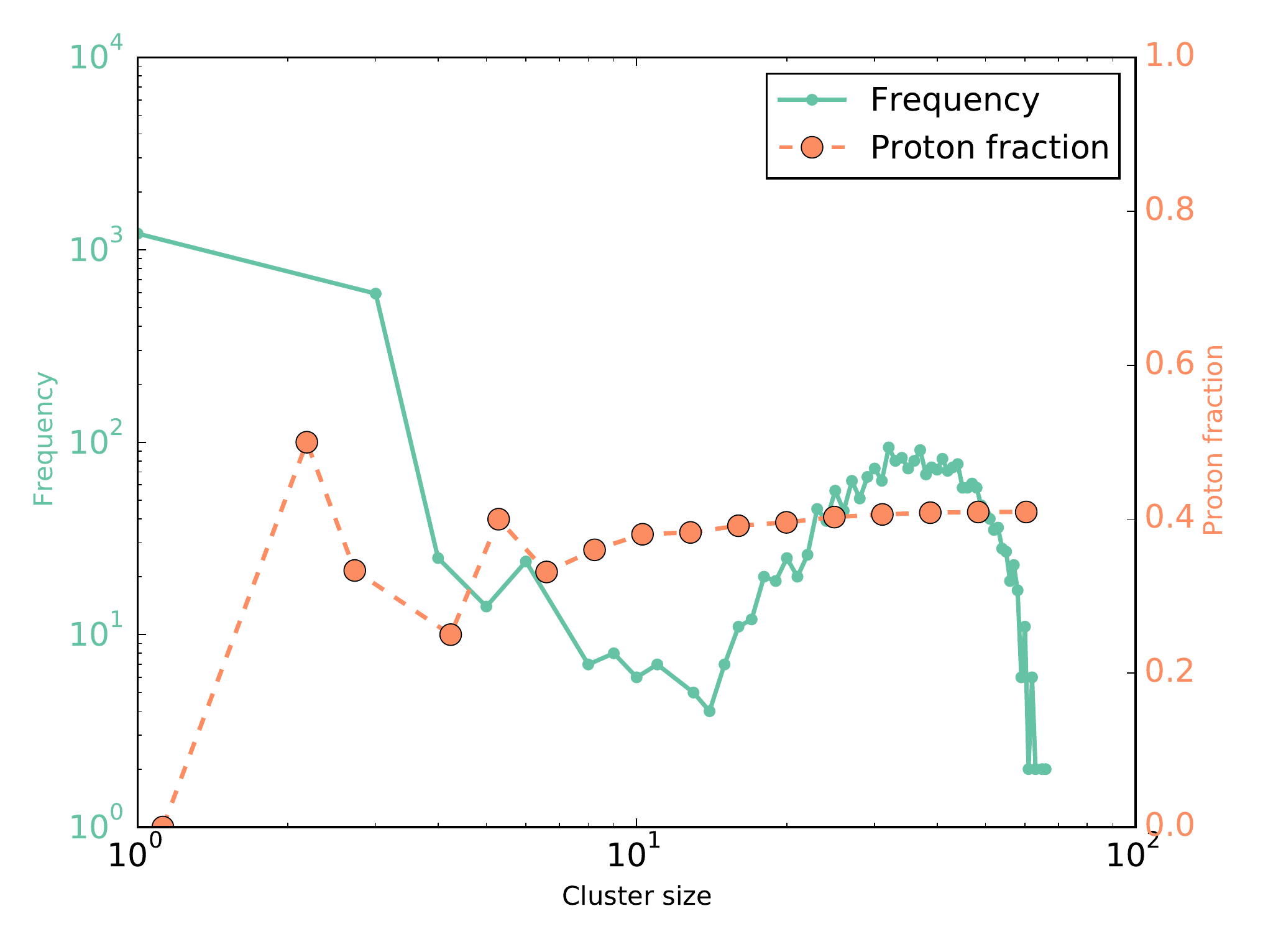}
    \caption{$\eta = 0.0001\,\text{fm/c}$}
    \label{subfig:1e-4}
  \end{subfigure}
  \begin{subfigure}[h!]{0.9\columnwidth}
    \includegraphics[width=\columnwidth]{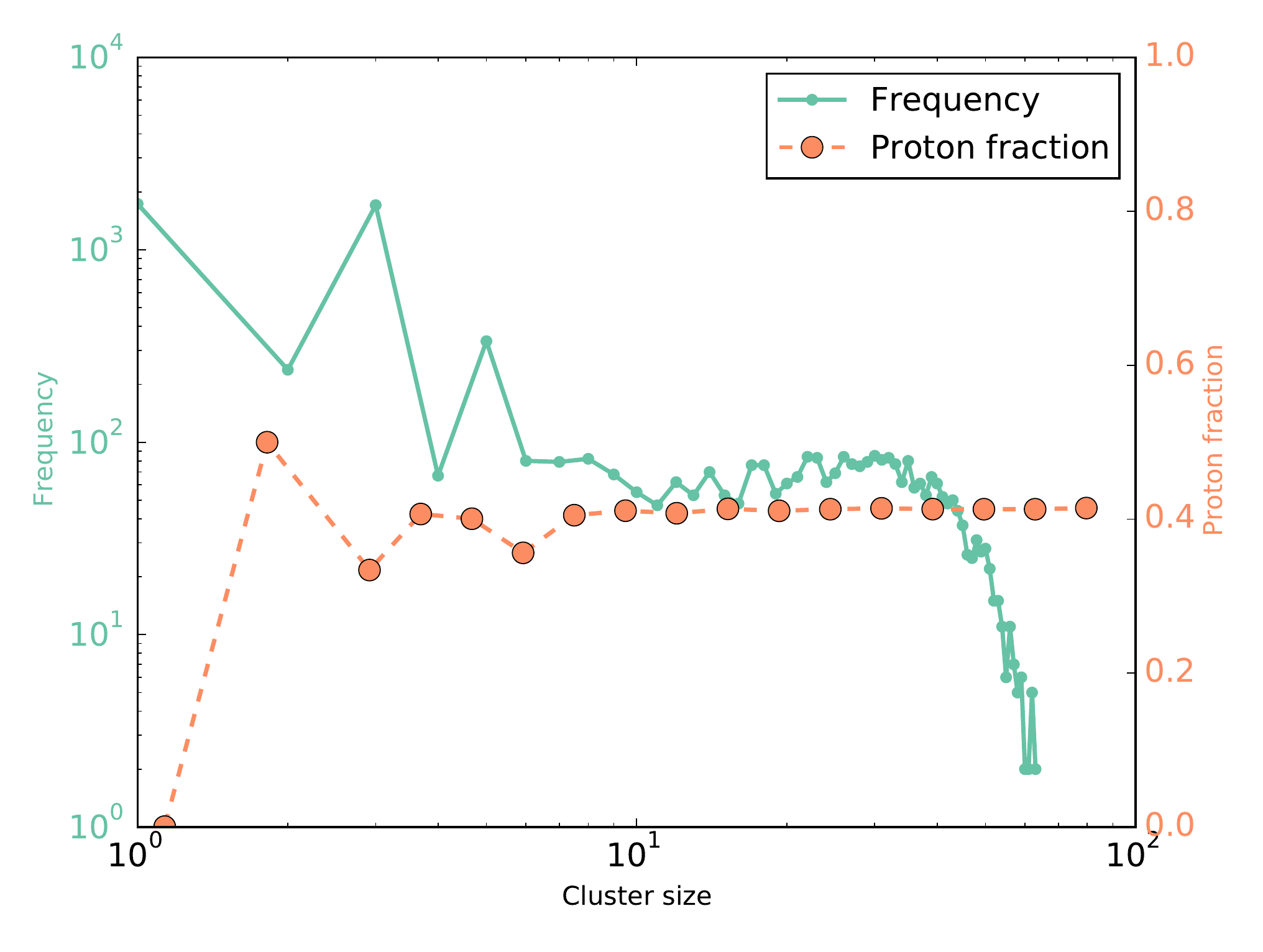}
    \caption{$\eta = 0.0005\,\text{fm/c}$}
    \label{subfig:5e-4}
  \end{subfigure}
  \begin{subfigure}[h!]{0.9\columnwidth}
    \includegraphics[width=\columnwidth]{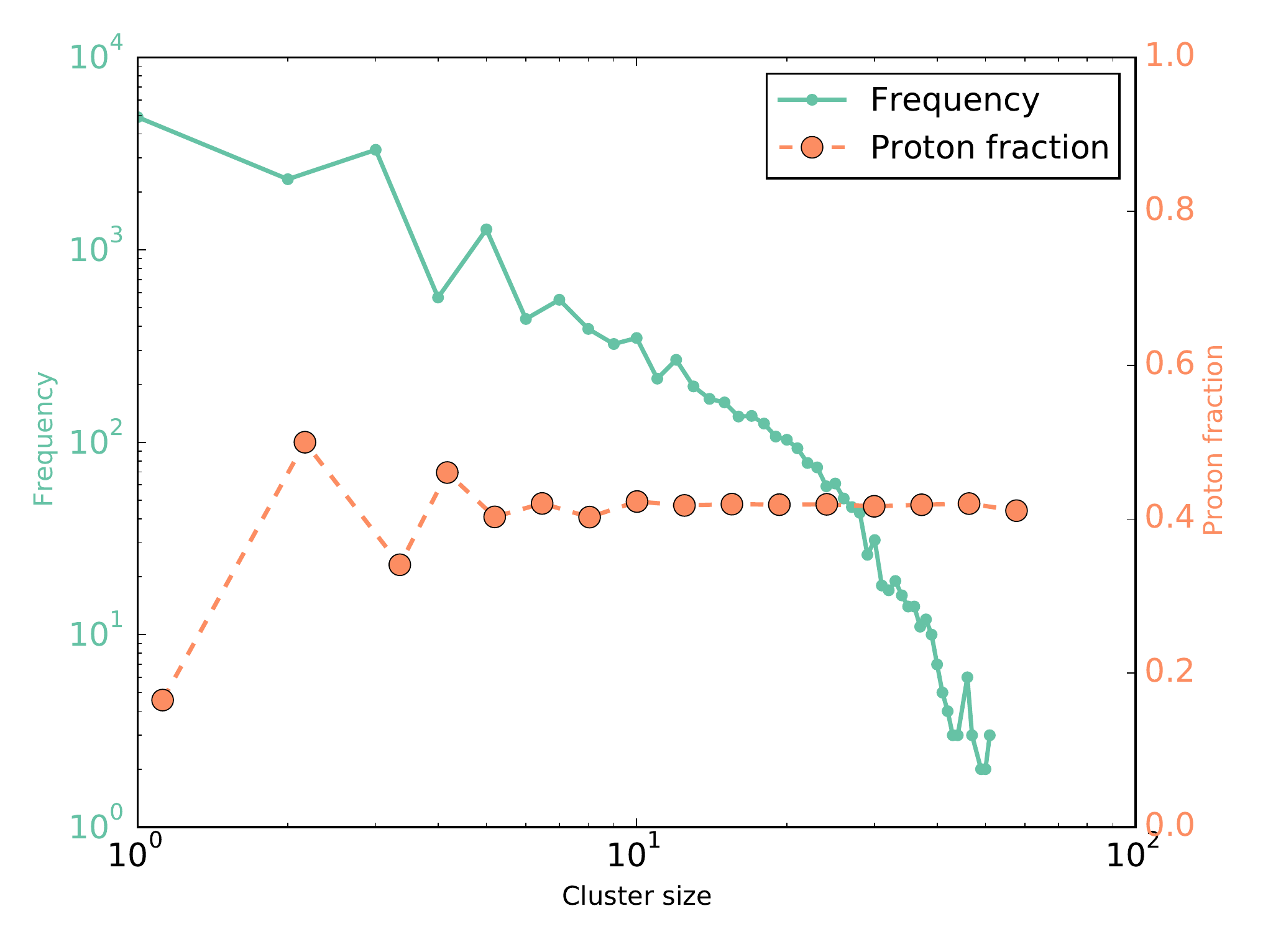}
    \caption{$\eta = 0.009\,\text{fm/c}$}
    \label{subfig:9e-3}
  \end{subfigure}
  \begin{subfigure}[h!]{0.9\columnwidth}
    \includegraphics[width=\columnwidth]{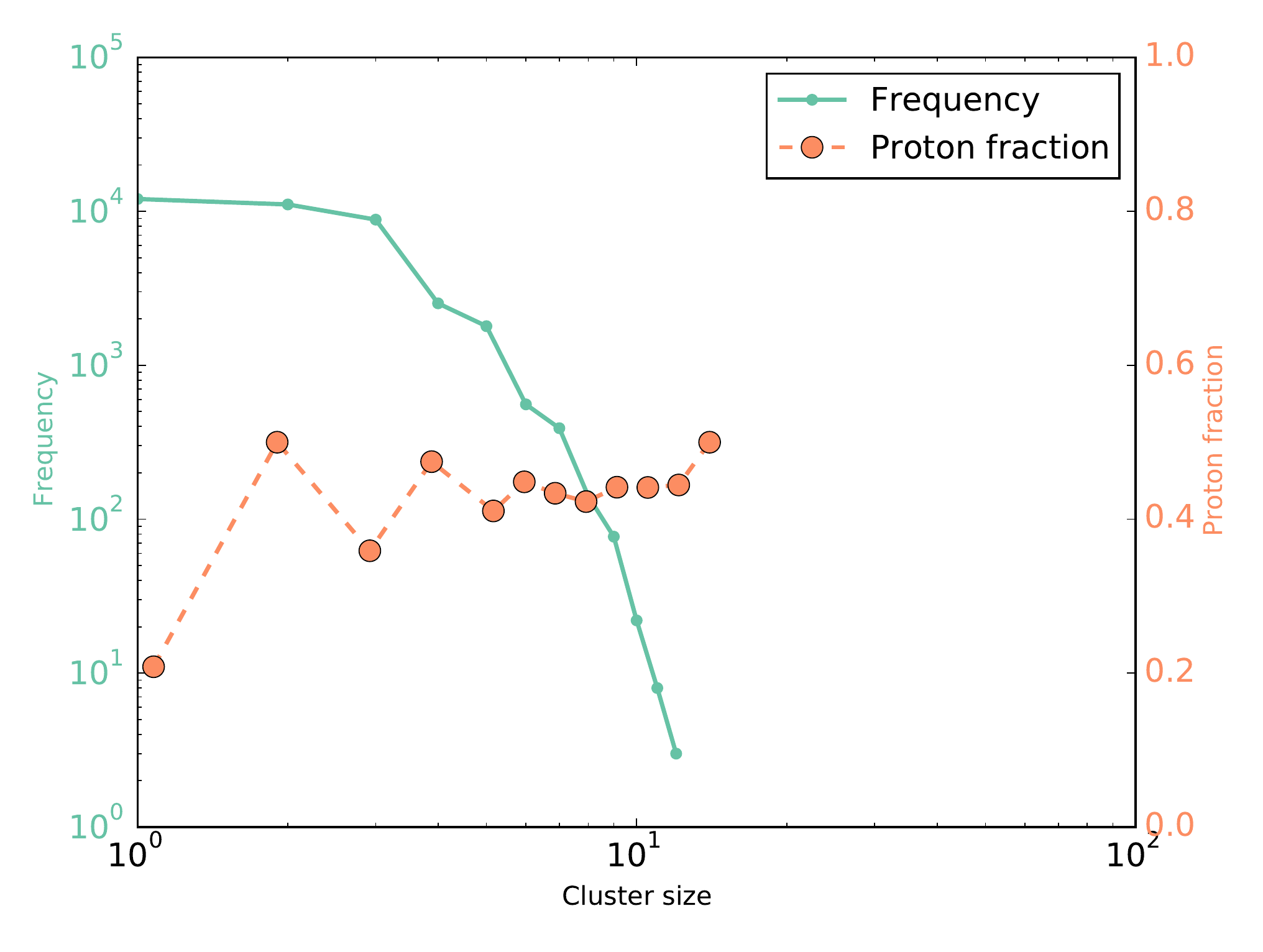}
    \caption{$\eta = 0.03\,\text{fm/c}$}
    \label{subfig:3e-2}
  \end{subfigure}
  \caption{(Color online) Fragment mass distribution. Labels grow with larger
    expansion rates. System consisting of 11000 particles with
    $x=0.4$ and initial density $\rho_0 = 0.08\,\text{fm}^{-3}$}
  \label{fig:distribution}
\end{figure*}

Figure~\ref{fig:distribution} shows the asymptotic fragment mass
distribution for 4 different expansion rates. In this case, the MSTE
algorithm has been applied over the primordial cell, taking into
account periodic boundary conditions knowing that there is no infinite
fragment as shown in figure~\ref{fig:infinite}. It can be seen that as
the homogeneous expansion velocity increases, the fragment mass
distribution displays the familiar transition from U-shaped to
exponential decay. Somewhere in between, a power law distribution is
to be expected. In particular, figure~\ref{subfig:9e-3} shows that
with an expansion of $\eta = 0.009\,\text{fm/c}$ we are close to a
power law distribution. It is interesting to note that at variance
with percolation or Lennard-Jones systems, due to the presence of the
Coulomb long range repulsion term, it is not possible to see an
infinite cluster in the asymptotic regime. Moreover, together with
the rather large rate of expansion, we do not expect to find big
clusters.

\section{Discussion and Concluding Remarks}

In this work we have performed numerical experiments of homogeneously
expanding systems with not very small amount of particles in the
primordial cell. In order to analyze the fragment structure of such a
system as a function of time, we have developed a graph-based tool for
the identification of infinite fragments for any definition of
percolation-like (i. e., additive) clusters. Once this formalism is
applied to the above mentioned simulations, we have been able to
identify the region in which a power-law distribution of masses is
expected. The fragment mass distribution shapes range from U-like to
exponential decay.

We are currently performing simulations with a larger number of
particles to properly characterize the critical behavior of the
system.

\bibliography{nuclear}
\end{document}